# A Phonon Boltzmann Study of Microscale Thermal Transport in α-RDX Cook-Off

Francis G. VanGessel, Gaurav Kumar, Daniel C. Elton and Peter W. Chung

Department of Mechanical Engineering
University of Maryland at College Park, College Park, MD, USA

**Abstract.** The microscale thermal transport properties of αRDX are believed to be major factors in the initiation process. In this study we present a thorough examination of phonon properties which dominate energy storage and transport in αRDX. The phonon lifetimes are determined for all phonon branches, revealing the characteristic time scale of energy transfer amongst phonon modes. The phonon parameters also serve as inputs to a Full Brillouin zone three dimensional phonon transport simulation in the presence of a hotspot. In addition to identifying the phonon mode contributions to thermal transport. As N-N bond breaking is integral to disassociation we identify phonon modes corresponding to large N-N bond stretch analyzing the manner in which these modes store and transfer energy.

**Introduction**

Phonons play an important role in overcoming energy barriers and initiating the chemical decomposition that can lead to detonation in secondary explosives such as αRDX [1, 2, 3]. However, open questions remain as to the exact nature of phonon-mediated energy transfer into the key internal molecular vibrational modes that result in chemical events. Dlott and coworkers have posited that this process occurs in an indirect manner through multiphonon up pumping [1]. However, other works speculated that the energy transfer process occurs through a direct route without intermediate energy transfer [4]. Thus, further investigation is needed to elucidate the manner in which phonons store, transport, and transfer energy.

Among the candidate approaches for investigating phonon behavior in general materials are phonon Boltzmann and molecular dynamics (MD) methods. Advances in methods as well as computer hardware capabilities have led to a surge of interest in out-of-equilibrium phonon transport mechanisms as well as determination of important phonon properties such as phonon mode lifetimes [5, 6]. Though MD still remains a widely used method, and despite the many developments that enable the study of larger material domains, such studies are still relatively limited in terms of length scales. Phonon Boltzmann is inherently a microscopic theory that forgoes some of the atomistic information in favor of reaching length scales more commonly seen in real materials. To date, however, phonon Boltzmann studies have been limited mainly to crystals with a relatively few number of atoms in the unit cell, likely due in part to the computational complexity associated with modelling phonons in materials whose unit cells contain large numbers of atoms. For instance, αRDX is a complex molecular crystal containing 168 atoms in its unit cell, resulting in 504 phonon branches in the first Brillouin zone (BZ). In the continuum limit, each branch is a continuous curve representing the infinite number of possible wave directions that may participate in the storage or transport of energy. Development of these approaches for energetics could reveal much about the behavior of thermal and vibrational energy as it flows and scatters within a highly heterogeneous microstructural mixture. Such developments have not been attempted, as far as we are aware.

In this work, we present recent results in our attempts to investigate the possible mechanisms by which energy is transferred from localized heating into the molecular vibrations of αRDX. We begin by determining the phonon thermal properties based on a complete and exhaustive representation of every phonon vibrational mode. We identify phonon modes associated with large N-N bond stretch, as scission of the N-N bonds is understood to be of particular importance to the disassociation of the RDX molecule [7]. The phonon properties are then used as inputs to a phonon Boltzmann simulation of localized heating, i.e. a hotspot, in an αRDX grain to study out-of-equilibrium thermal transport. Finally, we apply normal mode decomposition techniques to determine accurate phonon mode lifetimes. These lifetimes reveal important information regarding the time scales at which phonon modes scatter, a crucial parameter for understanding the viability of models such as this to study initiation.

**Methods**

In order to form a more complete understanding of microscale energy transport in αRDX we use a variety of methods. These methods are used to a) perform a full BZ analysis of the phonon vibrational modes present in αRDX, focusing on the thermal transport properties of the phonons as well as a normal mode analysis, b) calculate the full-band highly accurate phonon lifetime data within the single mode relaxation approximation for all 504 branches in RDX, c) perform phonon Boltzmann transport equation (BTE) simulation of a localized heating to investigate non-equilibrium energy transport in the presence of a hot spot, and d) identify phonon modes which cause large relative displacements between bonded Nitrogen atoms.

Phonon Thermal Properties

Phonon thermal properties are calculated using the intra and intermolecular interactions parameterized by the Smith and Bharadwaj potential [8]. Each phonon mode is defined by a unique wavevector, $\mathbf{k}$, and branch, $\lambda$. Namely, $\phi = (\mathbf{k}, \lambda)$. A study of phonon-mediated thermal flow in αRDX requires a number of phonon properties, each of which may span several orders of magnitude in value. These include frequency, $\omega_\phi$, specific heat, $C_\phi$, group velocity, $\mathbf{v}_\phi$, and relaxation time, $\tau_\phi$. With the exception of phonon relaxation time, calculation of these properties may be accomplished via LD using the software package GULP [9]. The LD approach allows for the determination not only of the phonon properties but also the phonon mode shapes [10].

Within the phonon gas model framework the thermal conductivity (TC) of a material can be expressed as [11]
$$\kappa_{ij} = \sum_\phi C_\phi v_{\phi i} v_{\phi j} \tau_\phi \qquad (1)$$
where $i, j \in (1,2,3)$ are spatial indices. For the initial BZ analysis, we employ a gray approximation for the phonon relaxation times assuming $\tau_\phi = 217$ ps for all phonon modes. This number is chosen to ensure that the average TC predicted by Eq. 1 agrees with the average TC published in [12], $\bar{\kappa} = 0.355$ J/mK.

In addition to the absolute TC we also consider the thermal conductivity accumulation (TCA) with respect to frequency,
$$\kappa_{ii}^{accum}(\omega) = \frac{1}{\kappa_{ii}} \sum_{\{\phi : \omega_\phi < \omega\}} C_\phi v_{\phi i}^2 \tau_\phi \qquad (2)$$
The TCA quantifies the relative contributions of different regions of the frequency spectrum to thermal transport.

Phonon Relaxation Times

Despite the relative importance of phonons to initiation mechanisms in energetics, surprisingly little is known about their basic anharmonic phonon properties like relaxation times. Accurate prediction of certain properties like TC require knowledge of anharmonic vibrational properties, particularly the relaxation times. Simple models of relaxation times based on Callaway's initial work [13] have shown good agreement with experimental TC values many materials. One result of Callaway's model is an empirical formula for the phonon relaxation time that assumes phonons relax to equilibrium at time scales inversely proportional to $\omega_\phi^f$, where the exponent $f$ is a fitted parameter. However, for many materials Callaway's model leads to inaccurate prediction of phonon relaxation times which may be attributed to use of additional fitting parameters [14]. Thus we require a more accurate description of phonon behavior than is provided by these empirical models. We remark that this part of the effort extends, for the first time, existing techniques previously only applied to atomic crystals to obtain accurate phonon relaxation times for *every* phonon branch in αRDX.

We use the normal mode decomposition (NMD) method to determine the phonon lifetimes which has been used numerous materials including crystalline silicon, disordered silicon, and argon [15]. For this study we use the same atomic potential as in previous sections. We consider a relatively small supercell consisting of $N = 2 \times 2 \times 2 = 8$ unit cells of $\alpha$RDX, whose structure has been energy-minimized in GULP using a constant pressure condition. Lattice dynamics is then used to obtain the phonon mode eigenvectors $e_{\phi,\alpha,\beta}$, where $\beta$ is basis index for an atom in the unit cell and $\alpha \in (1,2,3)$ is the Cartesian index. Next, a molecular dynamics (MD) trajectory is generated using LAMMPS [16] with 1 fs time-steps. The system is equilibrated initially for 1000 ps at 300K under NPT conditions. This was followed by a production step, using NVE conditions, during which displacement and velocity data were gathered every 0.004 ps for 2000.0 ps. From these quantities we calculate the phonon normal mode coordinate, $q_\phi(t)$, using

$$q_\phi(t) = \sum_{\alpha,\beta,l}^{3,n,N} \sqrt{\frac{m_\beta}{N}} u_{\alpha\beta l}(t) \, e^*_{\phi\alpha\beta} \, exp(i\mathbf{k} \cdot \mathbf{r_{ol}}) \quad (3)$$

where $u_{\alpha\beta l}$ represents the displacement from equilibrium in the $\alpha^{th}$ direction of the $\beta^{th}$ atom in the basis attached to the $l^{th}$ unit cell. $m_\beta$ is the mass of $\beta^{th}$ atom in the unit cell and $\mathbf{r_{ol}}$ is the equilibrium position vector of the $l^{th}$ unit cell. We determine $\dot{q}_\phi(t)$ by taking the time derivative. The mode projected SED is then defined as [6]

$$\Phi(\mathbf{k},\omega) = \left| \frac{1}{\sqrt{2\pi}} \int_{-\infty}^{\infty} \dot{q}_\phi \, exp(-i\omega t) \, dt \right|^2 \quad (7)$$

This is the power spectrum of the mode velocity, and reflects the kinetic energy density of each mode at different frequencies. In our case, we calculate SEDs separately for each mode. Each SED contains a single peak which was fit to a Lorentzian function under the ansatz $\Phi(\mathbf{k},\omega) = \sum_\nu C_{0\nu} \Gamma_\phi / [\pi(\omega_{0\phi} - \omega)^2 + \pi\Gamma_\phi^2]$ where $C_{o\phi} = \sum_j \sum_{j'} [cos(\omega(t_j - t_{j'}) A_{j\phi} \, A_{j'\phi} \frac{\omega_{o\phi} + \Gamma_\phi^2}{8 \, \tau_\phi \Gamma_\phi}$. Finally, the phonon relaxation time may be extracted from the Lorentzian line width $\Gamma_\phi$ as [15] $\tau_\phi = 1/(2\Gamma_\phi)$. The actual fitting of the phonon SED data was performed by fitting, in least squares sense, of the following Lorentzian function,

$$\Phi = \sqrt{P_3} \left[ P_1 \frac{P_2^2 + P_3}{4\pi} \right] / [(P_2 - \omega)^2 + P_3] \quad (4)$$

where $P_1, P_2, P_3$ are fitting parameters.

Phonon Boltzmann Transport Modelling

In order to model phonon behavior under nonequilibrium conditions, such as phonons in a hotspot, we use the full BZ three dimensional phonon BTE solution method detailed in [17]. Briefly, this method involves the solution of the phonon BTE in the relaxation time approximation (RTA), which is written

$$\frac{\partial e_\phi(r)}{\partial t} + \mathbf{v}_\phi \cdot \nabla e_\phi(\mathbf{r}) = \frac{e_\phi^0(T) - e_\phi(\mathbf{r})}{\tau_\phi} \quad (5)$$

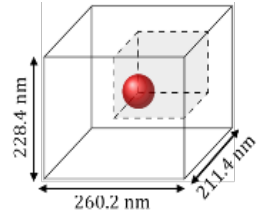

Figure 1: Physical domain representing a hotspot in RDX. The sphere indicates the hotspot region. Due to the symmetry, the simulation domain reduces to the shaded octant.

where $e_\phi$ is the phonon mode energy and $e_\phi^0(T) = [exp(\hbar\omega_\phi/k_B T) - 1]^{-1}$ is the equilibrium mode energy where $T = 300$ K. For simplicity, we initially consider the steady state problem and hold the hot spot in a fixed location. The solution yields the spatial variation of the modal energy values as well as the modal fluxes, $\mathbf{f}_\phi(\mathbf{r}) = \mathbf{v}_\phi e_\phi(\mathbf{r})$. The total energy and total flux values are calculated as a sum over the sampled phonon modes, i.e. $e(\mathbf{r}) = \sum_\phi e_\phi(\mathbf{r})$ and $\mathbf{f}(\mathbf{r}) = \sum_\phi \mathbf{f}_\phi(\mathbf{r})$. The phonon properties used here are the same as those obtained for analysis of the phonon thermal properties where a sampling of 172,368 phonon modes was used. Different, however, are the phonon relaxation times, $\tau_\phi$, which are obtained now using the NMD procedure. Due to the large computational requirements associated with obtaining $\tau_\phi$ throughout the BZ we use the relaxation times obtained at the BZ center and assume that $\tau_\phi$ is constant within each phonon branch. This therefore assumes that *scattering* is isotropic presently but distinct for each polarization. The steady state form of Eq. 3 is solved using the control volume method. We choose a brick-shaped domain corresponding to 200x200x200 unit cell tiles along the x, y, and z directions, 10,400 control volumes are used. A spherical hotspot with 30 nm radius is placed at the center of the domain. Within the hotspot region, $1 \times 10^{-10}$ J of energy is sourced into the acoustic modes. The physical domain is pictured in Figure 4 but due to symmetry we simulate only an octant of this domain. Symmetry planes are modeled as specular boundaries while the other boundaries allow phonons to flow freely out of the domain.

Bond-Stretch Metric

The stretching of N-N bonds is believed to play a fundamental role in the disassociation of the RDX molecule [18, 7]. Therefore, we wish to identify phonon modes that correspond to large N-N bond stretch as these modes may be a key factor in initiation. The perturbing effect of phonon mode $\phi$ causes atom $i$ to be displaced to the spatial location, $\boldsymbol{r}_{\phi i} = \boldsymbol{x}_i + \boldsymbol{u}_{\phi i}$. Here $\boldsymbol{x}_i$ is the equilibrium atomic location while the displacement due to the phonon is given by the vector $\boldsymbol{u}_{\phi i} = A_{\phi i} \boldsymbol{e}_{\phi i}$, where $\boldsymbol{e}_{\phi i}$ represents the components of the mode shape eigenvector $\boldsymbol{e}_\phi$ corresponding to atom $i$, and $A_{\phi i} = \sqrt{e_\phi/m_i \omega_\phi^2}$ is the phonon mode amplitude [10]. Therefore, given the energy in a phonon mode, the resulting spatial location can be calculated. Then, the $b^{\text{th}}$ N-N bond connecting atoms $b_1$ and $b_2$ is stretched to the length $\|\boldsymbol{r}_{\phi b_1} - \boldsymbol{r}_{\phi b_2}\|$. The bond stretch metric for mode $\phi$ is the maximum bond stretch length over all bonds,

$$\Delta_\phi = \max_b \|\boldsymbol{r}_{\phi b_1} - \boldsymbol{r}_{\phi b_2}\| \qquad (6)$$

In order to calculate the bond stretch we require values for the phonon energy of each individual mode. Here we consider two limiting cases. The first case assumes thermal equilibrium in which energy, and therefore amplitude, is determined by Bose statistics at temperature $T$. The second case corresponds to a non-equilibrium situation in which the modal amplitudes are uniformly constant. The second case allow us to consider modes that may not typically contain a large amount of energy, when in equilibrium, but whose displacements correspond to large N-N stretch. In the equilibrium case $A_{\phi i} = \sqrt{e_\phi^0/m_i \omega_\phi^2}$ while in the uniform case $A_{\phi i} = A_0$. Thus we introduce the modal N-N stretch metric, written $\Delta_\phi^{eq}$ and $\Delta_\phi^{uni}$ respectively, for the equilibrium (EQ) and uniform (UNI) cases respectively.

**Results**
Full Brillouin Zone Analysis of αRDX

The phonon frequency surfaces for a quadrant in the $k_z = 0$ plane of the RDX BZ is shown in Fig. 2., for clarity only the first 14 of the 504 total branches are shown. For each branch, we use a uniform sampling of 342 wavevectors throughout the BZ which results in 172,368 phonon modes. Figs. 3 and 3 show the contribution of discrete frequency intervals to the group velocity component magnitudes and specific heat respectively. The contribution from discrete frequency intervals to a thermal property is calculated using the equation

$$a(\omega) = \sum_{\{\phi : \omega_\phi \in [\omega, \omega + \Delta\omega]\}} a_\phi \qquad (7)$$

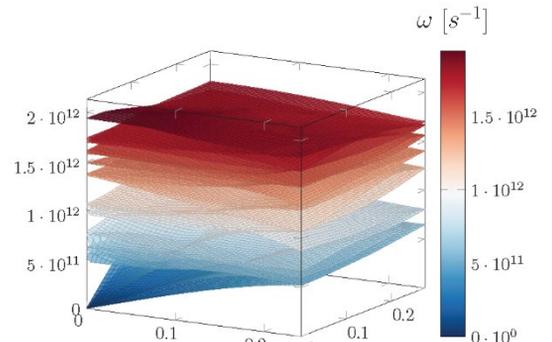

Figure 2: Frequency surfaces of first 14 branches in the quadrant of the $k_z = 0$ plane of the BZ.

where $a$ is any phonon mode property. The specific heat represents the phonon mode's ability to store energy, while the group velocity is a measure of the speed with which a phonon mode transports energy. The total specific heat is calculated to be $C_v = 4.11 \times 10^3$ kJ/m³ K which is roughly a factor of two of the experimental value $1.96 \times 10^3$ kJ/m³ K [12]. The discrepancy between our calculated value and the reported experimental value is due to the implicit assumption of 0K in LD. The inclusion of temperature would cause an overall reduction in the frequency values, reducing the specific heat value. Furthermore, the calculated group velocity magnitudes agree with those reported in [4], which is to be expected since the same atomic potential is presently used. Figure 4 indicates that the phonon frequency spectrum of RDX contains large band gaps punctuated by narrow bands. Additionally, in Fig. 3 we can see that phonon modes with the largest group velocities are located at low frequencies, but a few bands at higher frequencies do possess non-negligible group velocities. Therefore as noted in [4], we expect the optical modes to contribute significantly to thermal transport in RDX.

Calculating the TC tensor via Eq. 1 yields
$$\kappa_{xx} = 0.33 \frac{J}{mK}, \kappa_{yy} = 0.28 \frac{J}{mK}, \kappa_{zz} = 0.45 \frac{J}{mK}, \quad (8)$$
while all other components are negligible. Note that there exists a high degree of anisotropy in the thermal transport in αRDX with TC along the $z$ direction 1.6 times larger than along the $y$ direction. While the average TC calculated here matches with the published values in [12], our TC more anisotropic due to the gray phonon approximation. The calculation enables the consideration of anisotropy by sampling the phonons from the full three-dimensional BZ. The TCA is shown in Fig. 5 where we see that the phonons in the low frequency regime, $\omega < 1$ THz, contribute to a larger percentage of the thermal transport along the $y$ and $z$ directions, 20% and 25% respectively, than to thermal transport along the $x$ direction 15%. However, phonons within the frequency interval 2 THz $< \omega <$ 3 THz contribute roughly 24% of the total TC along the x-direction, causing $\kappa_{xx}$ to overtake $\kappa_{yy}$ and $\kappa_{zz}$ in this interval. Finally, note that phonons with frequencies $> 3 \times 10^{12}$ THZ contribute ~25% of the TC along the $y$ direction versus ~17.5% to conductivity along the $x$ or $z$ directions. We suspect that because phonon modes with frequencies > 10 THz contribute < 20% of the thermal conductivity, once energy is introduced to those modes, heat transfer will be mitigated, possibly leading to thermal runaway and ultimately cookoff.

The final portion of the BZ analysis focuses on the phonon modes that correspond to large stretch of N-N bonds. The bond stretch metric for the EQ, and UNI cases was detailed in the methods section.

Histograms of the mode stretches for the EQ and UNI cases are shown in Fig. 6. Both distributions are heavily weighted toward zero indicating the vast majority of modes contribute

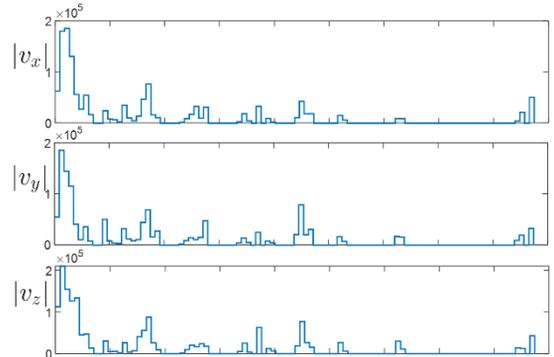

Figure 3: Stair plot of the group velocity component magnitudes vs. frequency.

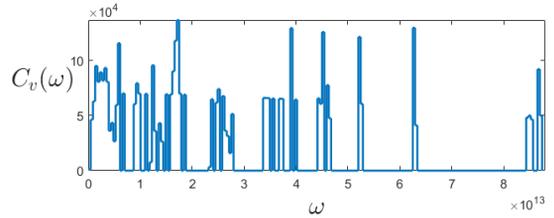

Figure 4: Stair plot of the specific heat vs. frequency.

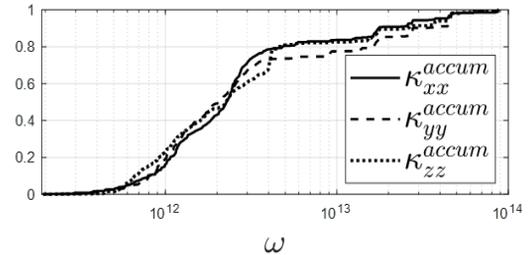

Figure 5: TCA vs. frequency along the three Cartesian directions.

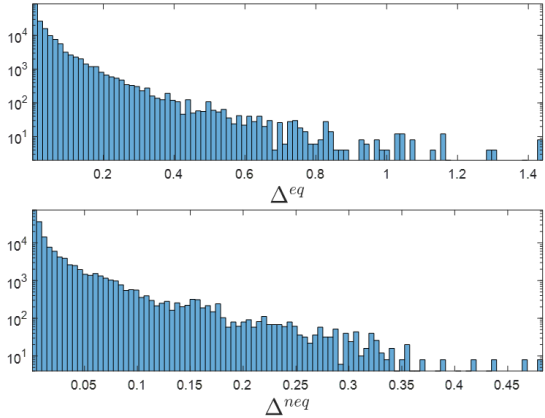

Figure 6: Histogram of phonon mode stretches for the EQ (top) and UNI (bottom) case.

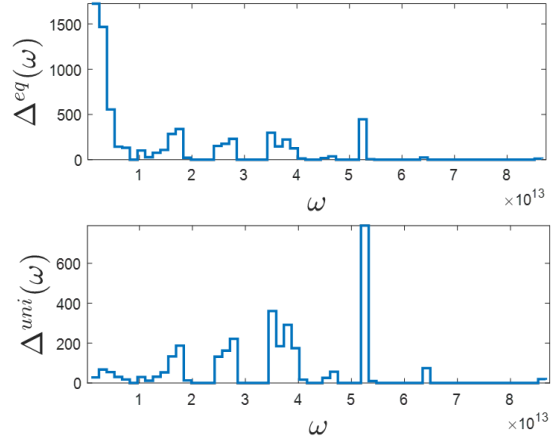

Figure 7: Plot of N-N bond stretch contribution from discrete intervals in the frequency spectrum. The bond stretch has units of angstroms.

negligibly to N-N bond stretching. The contributions of phonons within discrete frequency intervals to the EQ and UNI mode stretch is calculated using Eq. 7 with $a_\phi = \Delta_\phi^{eq}, \Delta_\phi^{uni}$ and is visualized in Fig. 7. We note that the frequency intervals corresponding to the largest N-N stretch differ between the EQ and UNI cases. Phonons with frequencies <10THz contributes significantly to the N-N bond stretch according to the EQ metric, but contributes negligibly using the UNI metric. However, frequency intervals around 17.5, 27.5, and 37.5 THz contribute to non-negligible bond stretching for both metrics. Furthermore, we observe a large peak in the bond stretches at 52.5 THz for both the EQ and UNI metrics. Presence of this peak in both measures of bond stretch indicates the frequency interval around 52.5 THz will play a major role in facilitating N-N bond scission. Thus using our bond stretch metric we have identified the region of the frequency spectrum most likely to be involved in initiation. Note that our prediction of 52.5 THz compares well with other work which predicted the important N-N bond stretching mode to be in the 30-60 THz range [1, 2]

Full-Band Relaxation Times of RDX

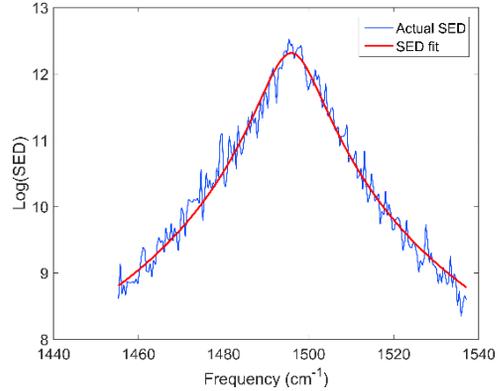

Figure 8: SED vs Frequency for an optical branch of 2x2x2 supercell.

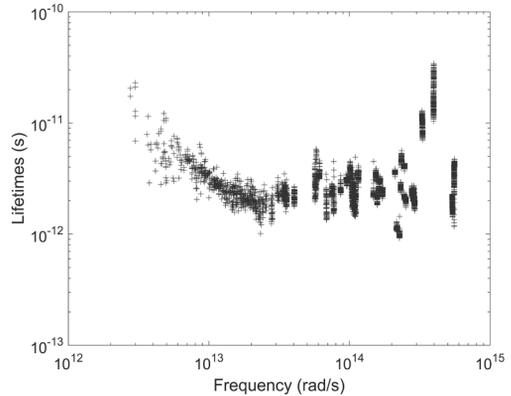

Figure 9: Phonon Lifetimes vs Frequency for RDX 2x2x2 supercell.

Shown in Fig. 8 is the Lorentzian fitting result for a single representative SED[a]. The following metric was used to quantify the error in SED-fitting

$$\text{error}_\phi = \sqrt{\frac{\sum_{l=1}^{n_d}(SED_{\phi l} - SEDfit_{\phi l})^2}{n_d}} \quad (9)$$

$$\% \text{ error} = 100 \frac{\sqrt{\frac{\sum_{\phi=1}^{3n} Error_\phi^2}{3n}}}{\sqrt{\frac{\sum_{\phi=1}^{3n}\sum_1^{n_d}(SED_\phi)^2}{3n \times n_d}}} \quad (10)$$

where $\text{error}_\phi$ is root mean square error in fitting mode $\phi$ and, $n_d$ is the number of data points in each SED mode, and $3n$ is the total number of sampled modes. Using the above metric we get 2.61% error in fitting SED for RDX 2x2x2. Standard deviation of $\text{error}_\phi$ values over all modes was calculated as

$$\sigma_{\text{fit}} = \frac{\sqrt{\frac{\sum_\phi^{3n}\left(error_\phi - \frac{\sum_\phi^{3n} error_\phi}{3n}\right)^2}{3n}}}{\frac{\sum_\phi^{3n} error_\phi}{3n}} \quad (11)$$

$\sigma_{\text{fit}}$ was found to be 0.006583 suggesting consistent good fit for all sampled modes. Figure 9 indicates that the phonon lifetime values for a 2x2x2 supercell of RDX fall in the range of 0.9217 ps to 34.1581 ps. Similar results were obtained for RDX 1x1x1. This is in good agreement with lifetime values for vibrons in RDX which have been reported to fall in the range of 2.5 ps to 11 ps [19]. Another source reported the normal stress and shear stress relaxation times, contributed by lattice vibration and molecular rotation respectively, for β-HMX to fall in range of 0.8347 ps to 27.4603 ps [20]. Further, the number of phonon modes was plotted against lifetime values and a bell-shaped curve was observed as shown in Fig. 10. It was observed that the majority of phonon modes are concentrated in region of small relaxation time values and there is little spread in the lifetimes data.

A substantial contribution to phonon TC comes from three acoustic modes, but their lifetime values cannot be defined at the gamma point. In addition, group velocities for many optical branch phonons vanish at the gamma point and edge of the BZ. Hence, a finer sampling in the BZ zone is needed for more accurate prediction of TC values. For RDX 1x1x1, the only allowed wave-vector is the gamma point, and for RDX 2x2x2, the allowed wave vectors lie at the gamma point and edge of BZ. This comes from imposing periodic boundary condition at material boundary [21]. Computation of phonon properties with a larger supercell for better sampling of the BZ is under progress.

Phonon BTE Simulation of RDX Hotspot

In this section we present the results of a full BZ three-dimensional phonon BTE simulation of localized heating, i.e. hotspot, in αRDX. It is believed that hotspot formations play an integral role in exciting the intramolecular vibrations associated with N-N bond breaking, i.e. the large $\Delta_\phi$ modes [4, 2]. Previous studies have generally assumed that phonons within the hotspot are in equilibrium at the hotspot temperature [1]. Attempts to include non-equilibrium thermal effects have often assumed the hotspot energy is distributed only among low frequency vibrational modes [4, 1]. Here, we seek a quantitative approach for determining the actual distribution of out-of-equilibrium phonon mode occupations as they evolve in space. This will provide a more complete picture of how energy is distributed among phonon modes, which modes are responsible for the transport of thermal energy away from the hotspot, and what role modes corresponding to large $\Delta_\phi$ play.

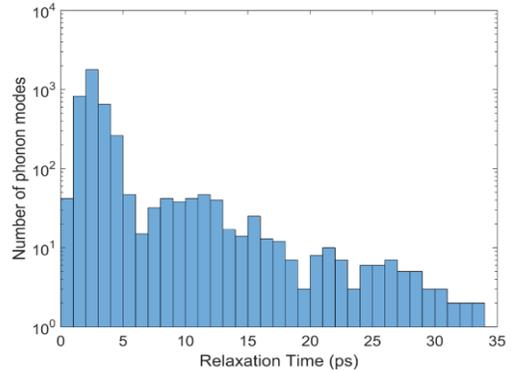

Figure 10: Number of phonon modes vs phonon relaxation time for RDX 2x2x2

---

[a] All 501 relaxation times and fitted curves are available upon request. Due to space limitations, they are not shown here.

We analyze the spatial variation of the phonon energy along the $x$, $y$, and $z$ axes emanating from the center of the hotspot. The absolute phonon energy along each direction is plotted in Fig. 11. The total energy rapidly decreases away from the hotspot center, reaching <1% of the maximum energy value when $r > 75$ nm. In αRDX the thermal carriers are unable to efficiently remove heat from the hotspot region. In addition, the discrepancy of the energy values along the $x$, $y$, and $z$ direction for the same $r$ values is indicative of the effect of lattice anisotropy on the TC.

To better understand the importance that modes with large N-N bond stretch might have, let us now choose the first 1724 modes that have the largest N-N bond stretches. This is 1% of the 172,368 total modes in the system. We define two different sets corresponding to the EQ and UNI conditions, $\delta^{eq}$ and $\delta^{uni}$ respectively. The fractional energy residing in $\delta^{eq}$ (and similarly for $\delta^{uni}$), calculated as $1/e_{tot} \times \sum_{\phi \in \delta^{eq}} e_\phi$ where $e_{tot} = \sum_\phi e_\phi$, is reported in Table 1. Incidentally, we note that the intersection of these sets is empty and the energy in each set is roughly equal to 1%. The percentage of energy in $\delta^{eq}$ is slightly larger than in $\delta^{uni}$ likely because modes in $\delta^{uni}$ have higher frequencies, which are generally less populated than lower frequency modes.

Finally, we analyze the modal thermal flux behavior in the presence of a hotspot.

Table 1: Phonon energy in stretching modes.

| % Energy Residing in Stretching Modes | |
|---|---|
| $\delta^{eq}$ | 1.08% |
| $\delta^{uni}$ | 0.94% |

The spatial variation in flux magnitude in the three Cartesian directions is plotted in Fig. 12, the flux reaches a maximum at the hot spot boundary regardless of direction. This indicates relatively more energy resides in modes with large group velocities in that spatial region. Furthermore, we note the significantly larger flux along the $z$-direction, as our BZ analysis predicted. Kraczek and Chung predicted that optical modes will carry a significant portion of the flux in αRDX [4]. We find this to be the case with the exact percentages reported in Table 2.

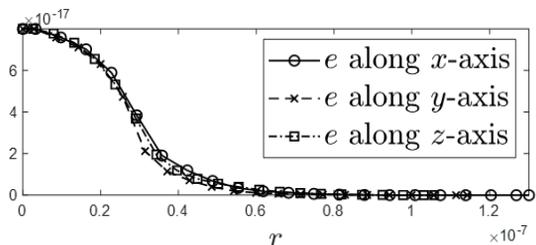

Figure 11: Total energy along three Cartesian directions emanating out from the hotspot center.

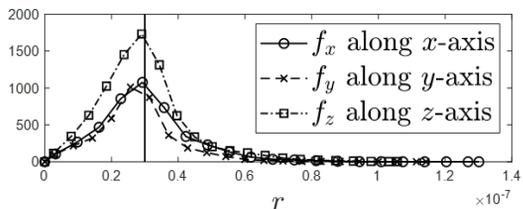

Figure 12: Flux components along the Cartesian directions emanating from hotspot center. The vertical black line indicates the hotspot boundary.

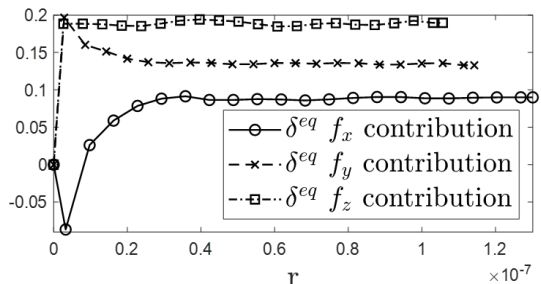

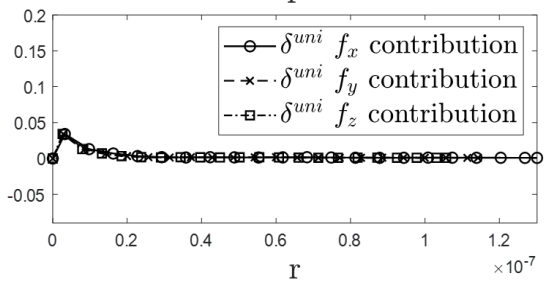

Figure 13: Contributions of EQ and UNI large stretch mode to the thermal flux along all three principal directions.

Table 2: Percentage of flux transported by acoustic and optical modes along principal directions.

|  | Acoustic | Optical |
|---|---|---|
| $f_x$ along $x$-axis | 26% | 74% |
| $f_y$ along $y$-axis | 22% | 78% |
| $f_z$ along $x$-axis | 37% | 63% |

Table 2 indicates a significant departure in the manner in which energy transfer occurs in αRDX compared to simple atomic crystals. Whereas optical mode contribution is minimal in atomic crystals, these modes are dominant contributors to the thermal flux in αRDX.

Finally, we analyze the flux contribution from the $\delta^{eq}$ and $\delta^{uni}$ stretching modes. The flux contribution from these modes is calculated for $\delta^{eq}$ (and similarly for $\delta^{uni}$) using the equation $1/f_i \times \sum_{\phi \in \delta^{eq}} f_{\phi i}$ where $f_i = \sum_\phi f_{\phi i}$ and $i = x, y, z$. The contribution to the thermal flux from both the $\delta^{eq}$ and $\delta^{uni}$ stretching modes along the three principal directions is plotted in Fig. 13. We observe that thermal flux contributions are anisotropic for the EQ class but isotropic for the UNI class. The EQ class transports energy preferentially along the $z$-direction, and to a lesser degree along the $y$ axis. Indeed the thermal flux along the $y$ axis exceeds that along the $z$ axis, this behavior is the reverse of the TC tensor in Eq. 8 where $\kappa_{xx} > \kappa_{yy}$. Note that $\delta^{eq}$ contributes negative flux near the hotspot center due to significant energy residing in $\delta^{eq}$ phonon modes which travel in the $-x$ direction, however the total flux, $f_x$, remains positive at this location. In addition, the overall magnitude of the flux contribution from $\delta^{eq}$ is a factor of 2-10 times larger than the $\delta^{uni}$ contribution as a result of the relatively larger group velocities of the phonon modes constituting $\delta^{eq}$. Furthermore, while the ratio of energy residing in $\delta^{eq}$ was nearly equal to the mode number fraction, $\delta^{eq}$ carries a *larger* proportion of the thermal flux with respect to the mode number fraction. While $\delta^{eq}$ corresponds to only 1% of the total number of modes it contributes 9-20% of the flux. In contrast, $\delta^{uni}$ contributes ~.1% of the total thermal flux outside the hotspot region. Thus, we surmise that the modes closely linked with N-N bond stretch are capable of storing thermal energy while the modes corresponding to $\delta^{eq}$ play a significant role in transporting energy from the hotspot.

**Conclusions**

The phonon parameters of αRDX have been determined throughout the entire BZ and their contributions to thermal transport analyzed. Significant anisotropy in the TC is reported, while the TCA indicates contribution to the TC from different phonons vary depending on the phonon frequency as well as direction of heat transport. We also present two metrics for determining the N-N bond stretch due to a certain phonon mode. These two metrics are based on whether the system is in equilibrium or non-equilibrium. The vast majority of modes contribute negligibly to N-N bonds stretch. We presented our initial findings for phonon lifetimes for all branches and at the BZ center as well as BZ boundary. The phonon lifetimes we found range from 1-40 ps. Finally we simulated microscale heat transport in the presence of a hotspot. We found that the optical modes contributed up to 75% of the total thermal flux. Furthermore, the energy stored in phonon modes corresponding to large N-N bond stretch was nearly equal to the number fraction of such modes. Additionally, we found that phonon modes with large N-N bond stretch under equilibrium conditions contribute significantly, 8-20%, to thermal transport, while stretching modes determined from the non-equilibrium assumption contributed negligibly to thermal flux. This work represents our first steps to applying accurate atomic and microscale thermal simulation techniques to more accurately quantify the phonon processes involved in cook-off of αRDX.

**Acknowledgements**

F. V. gratefully acknowledges the graduate fellowship from the Center for Engineering Concepts Development. This work was also supported, in part, by the Army Research Office under Award W911NF-14-1-0330 and the Department of Mechanical Engineering at the University of Maryland.

References

[1] D. D. Dlott and M. D. Fayer, "Shocked moelcular solids: Vibrational up pumping, defect hot spot formation, and the onset of chemistry," *The Journal of Chemical Physics,* vol. 92, no. 6, pp. 3798-3812, 1990.

[2] A. Tokmakoff, M. D. Fayer and D. D. Dlott, "Chemical Reaction Initiation and Hot-Spot Formation in Shocked Energetic Molecular Materials," *Journal of Physical Chemistry,* vol. 97, no. 9, pp. 1901-1913, 1993.

[3] S. Ye, K. Tonokura and M. Koshi, "Vibron dynamics in RDX, b-HMX and Tetryl crystals," *Chemical Physics,* vol. 293, no. 1, pp. 1-8, 2003.

[4] B. Kraczek and P. W. Chung, "Investigation of direct and indirect phonon-mediated bond excitation in α-RDX," *The Journal of Chemical Physics,* vol. 138, no. 7, p. 4505, 2013.

[5] F. G. VanGessel, J. Peng and P. W. Chung, "A review of computational phononics: the bulk, interfaces, and surfaces," *Journal of Material Science,* vol. 53, no. 8, pp. 5641-5683, 2017.

[6] J. M. Larkin, McGaughey and A. J. H, "Predicting alloy vibrational mode properties using lattice dynamics calculations, molecular dynamics simulations, and the virtual crystal approximation," *Journal of Applied Physics,* vol. 114, no. 2, p. 3507, 2013.

[7] I. V. Schweigert, "Ab Initio Molecular Dynamics of High-Temperature Unimolecular Dissociation of Gas-Phase RDX and Its Dissociation Products," *The Journal of Physical Chemistry A,* vol. 119, no. 1, pp. 2747-2759, 2015.

[8] G. D. Smith and R. K. Bharadwaj, "Quantum chemistry based force field for simulations of HMX," *Journal of Physical Chemistry,* vol. 103, no. 18, pp. 3570-3575, 1999.

[9] J. Gale and A. L. Rohl, "The General Utility Lattice Program (GULP)," *Molecular Simulation,* vol. 29, no. 5, pp. 291-341, 2003.

[10] B. T. M. Willis and A. W. Pryor, Thermal Vibrations in Crystallography, Cambridge: Cambridge University Press, 1975.

[11] P. G. Klemens, "Thermal conductivity and lattice vibrational modes," in *Solid state physics*, Elsevier, 1958, pp. 1-98.

[12] S. Izvekov, P. W. Chung and B. M. Rice, "Non-equilibrium molecular dynamics simulation study of heat transport in hexahydro-1,3,5-trinitro-s-triazine (RDX)," *International Journal of Heat and Mass Transfer,* vol. 54, no. 1, pp. 5623-5632, 2011.

[13] J. Callaway, "Model for Lattice Thermal Conductivity at Low Temperatures," *Phys. Rev.,* vol. 113, no. 4, pp. 1046-1051, 1959.

[14] A. J. H. McGaughey and M. Kaviany, "Quantitative validation of the Boltzmann transport equation phonon thermal conductivity model under the single-mode relaxation time approximation," *Phys. Rev. B,* vol. 69, no. 9, pp. 4303-4314, 2004.

[15] J. M. Larkin, "Vibrational Mode Properties of Disordered Solids from High-Performance Atomistic Simulations and Calculations Submitted in partial fulfillment of the requirements for the degree of Doctor of Philosophy in Mechanical Engineering.," Pittsburgh, 2013.

[16] S. Plimpton, "Fast Parallel Algorithms for Short-Range Molecular Dynamics," *Journal of Computational Physics,* vol. 117, no. 1, pp. 0021-9991, 1995.

[17] F. G. VanGessel and P. W. Chung, "An anisotropic full Brillouin zone model for the three dimensional phonon Boltzmann transport equation," *Computer Methods in Applied Mechanics and Engineering,* vol. 317, no. 1, pp. 1012-1036, 2017.

[18] J. E. Patterson, Z. A. Dreger, M. Miao and Y. M. Gupta, "Shock Wave Induced Decomposition of RDX: Time-Resolved Spectroscopy," *Journal of Chemical Physics,* vol. 112, no. 32, pp. 7374-7382, 2008.

[19] S. Ye and M. Koshi, "Theoretical Studies of Energy Transfer Rates of Secondary Explosives," *The Journal of Physical*

*Chemistry B,* vol. 110, no. 37, pp. 18515-18520, 2006.

[20] Y. Long and J. Chen, "A theoretical study of the stress relaxation in HMX on the picosecond time scale," *Modelling and Simulation in Materials Science and Engineering,* vol. 23, no. 8, pp. 5001-5023, 2015.

[21] C. Kittel, "Introduction to Solid State Physics," Eighth ed., John Wiley & Sons, Inc, 2005, pp. 105-127.

**Question from Dr. Joe Hooper**
Since the phonon scattering and IVR processes are generally picosecond scale, can you clarify how you would expect to achieve a significant phonon non-equilibrium during cookoff events which are vastly slower?

**Answer From Dr. Peter Chung**
One scenario is of an energetic material held at a fixed temperature that spontaneously initiates. In this case, the time scale of the cook-off may seem to be infinitely slow. However, the cook-off event is discrete. We hypothesize that the fluctuations associated with phonons that are constantly driving the system locally into and out of non-equilibrium are the keys to understanding this scenario of initiation. The local phonon fluctuations that can support a material's temperature are associated with a very fast time scale, far faster than the scales associated with thermal diffusion. However, this is still several orders of magnitude slower than the scale of atomic vibrations (femtoseconds).